\documentclass[twocolumn,english,prl,showpacs,aps,superscriptaddress]{revtex4-1}
\usepackage[T1]{fontenc}
\usepackage[latin9]{inputenc}
\usepackage{amsmath}
\usepackage{amssymb}
\usepackage{esint}
\usepackage{times}
\usepackage{graphicx}
\usepackage{color}

\newcommand{\vrec}{\ensuremath{v_\mathrm{rec}}}

\makeatletter
\@ifundefined{textcolor}{}
{%
 \definecolor{BLACK}{gray}{0}
 \definecolor{WHITE}{gray}{1}
 \definecolor{RED}{rgb}{1,0,0}
 \definecolor{GREEN}{rgb}{0,1,0}
 \definecolor{BLUE}{rgb}{0,0,1}
 \definecolor{CYAN}{cmyk}{1,0,0,0}
 \definecolor{MAGENTA}{cmyk}{0,1,0,0}
 \definecolor{YELLOW}{cmyk}{0,0,1,0}
 }

\@ifundefined{definecolor}

\usepackage{babel}

\makeatother

\begin{document}
\title{Violation of the Cauchy-Schwarz inequality with matter waves} 

\author{K. V. Kheruntsyan}
\affiliation{The University of Queensland, School of Mathematics and Physics, Brisbane, Qld 4072, Australia}
\author{J.-C.~Jaskula}
\altaffiliation{Current Address: Harvard-Smithsonian Center for Astrophysics, Cambridge, MA 02138.}
\affiliation{Laboratoire Charles Fabry de l'Institut d'Optique, CNRS, Univ Paris-Sud, Campus Polytechnique RD128, 91127 Palaiseau, France}
\author{P.~Deuar}
\affiliation{Institute of Physics, Polish Academy of Sciences, Al. Lotnik\'{o}w 32/46, 02-668 Warsaw, Poland}
\author{M.~Bonneau}
\affiliation{Laboratoire Charles Fabry de l'Institut d'Optique, CNRS, Universit\'{e} Paris-Sud, Campus Polytechnique RD128, 91127 Palaiseau, France}
\author{G.~B.~Partridge}
\altaffiliation{Current Address: Agilent Laboratories, Santa Clara, CA 95051.}
\affiliation{Laboratoire Charles Fabry de l'Institut d'Optique, CNRS, Universit\'{e} Paris-Sud, Campus Polytechnique RD128, 91127 Palaiseau, France}
\author{J.~Ruaudel}
\affiliation{Laboratoire Charles Fabry de l'Institut d'Optique, CNRS, Universit\'{e} Paris-Sud, Campus Polytechnique RD128, 91127 Palaiseau, France}
\author{R.~Lopes}
\affiliation{Laboratoire Charles Fabry de l'Institut d'Optique, CNRS, Universit\'{e} Paris-Sud, Campus Polytechnique RD128, 91127 Palaiseau, France}
\author{D.~Boiron}
\affiliation{Laboratoire Charles Fabry de l'Institut d'Optique, CNRS, Universit\'{e} Paris-Sud, Campus Polytechnique RD128, 91127 Palaiseau, France}
\author{C.~I.~Westbrook}
\affiliation{Laboratoire Charles Fabry de l'Institut d'Optique, CNRS, Universit\'{e} Paris-Sud, Campus Polytechnique RD128, 91127 Palaiseau, France}

\date{\today}

\begin{abstract}

The Cauchy-Schwarz (CS) inequality -- one of the most widely used and important inequalities in mathematics -- can be formulated as an upper bound to the strength of 
correlations 
between classically fluctuating quantities. 
Quantum-mechanical correlations can, however, exceed classical bounds.
Here we realize four-wave mixing of atomic matter waves using colliding Bose-Einstein condensates, 
and demonstrate the violation of a multimode 
CS inequality for atom number correlations 
in opposite zones of the collision halo.
The correlated atoms have large spatial separations and therefore 
open new opportunities for extending fundamental quantum-nonlocality tests to ensembles of massive particles.

\end{abstract}


\maketitle

The Cauchy-Schwarz (CS) inequality 
is ubiquitous in mathematics and physics \cite{CS-book}. 
Its utility ranges from proofs of basic theorems in linear algebra to the derivation of the Heisenberg uncertainty principle. In its basic form, 
the CS inequality simply states that the absolute value of the inner product of two vectors cannot be larger than the product of their lengths. 
In probability theory and classical physics the CS inequality can be applied to fluctuating quantities
and states that the 
expectation value of the cross-correlation $\langle I_1 I_2 \rangle$ between two quantities $I_1$ and $I_2$ 
is bounded from above by the auto-correlations in each quantity:
\begin{equation}
|\langle I_1I_2 \rangle| \leq \sqrt{\langle I_1^2 \rangle \langle I_2^2 \rangle}.
\label{cs1}
\end{equation} 
This inequality is satisfied, for example, by two classical currents 
emanating from a common source.

In quantum mechanics, correlations can, however, be stronger than
those allowed by the CS inequality \cite{Glauber:63,Reid:86,Walls:08}. 
Such correlations have been demonstrated in quantum optics using, for example, antibunched photons produced via spontaneous 
emission \cite{Kimble-Dagenais-Mandel-77}, or twin photon beams generated in a radiative cascade \cite{Clauser:74}, parametric down conversion \cite{Zou:91}, and optical four-wave mixing \cite{Marino:08}.
Here the discrete nature of the light and the strong correlation (or anticorrelation in antibunching) between photons 
is responsible for the violation of the CS inequality.
The violation has even been demonstrated for
two light beams detected as continuous variables~\cite{Marino:08}.

\begin{figure}[tb]
\includegraphics[width=8cm]{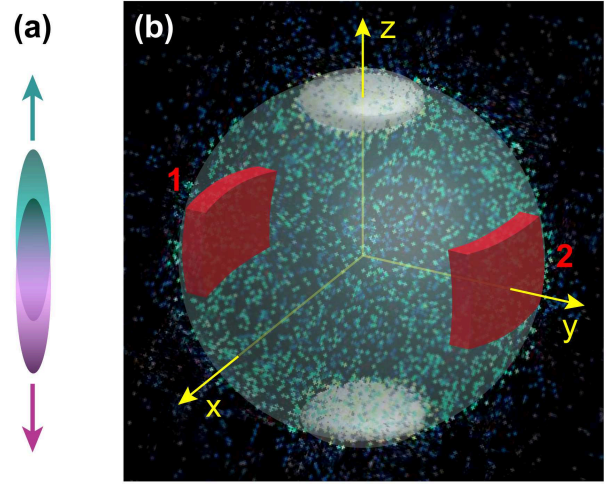}
\caption{(Color online) Diagram of the collision geometry. (a) Two cigar-shaped condensates moving in opposite directions along the axial direction $z$ shortly after their creation by a Bragg laser pulse (the anisotropy and spatial separation are not to scale). (b) Spherical halo of scattered atoms produced by four-wave mixing after the cloud expands and the atoms fall to the detector $46$ cm below. During the flight to the detector, the unscattered condensates acquire a disk shape shown in white on the north and south poles of the halo. The (red) boxes $1$ and $2$ illustrate a pair of diametrically symmetric counting zones (integration volumes) for the average cross- and auto-correlation functions, 
$\overline{\cal{G}}^{(2)}_{12}$ and $\overline{\cal{G}}^{(2)}_{ii}$ ($i=1,2$) (see text), 
used in the analysis of the Cauchy-Schwarz inequality.}
\label{fig:1}
\end{figure}

In this work we demonstrate a violation of the CS inequality in 
matter-wave optics using pair-correlated atoms formed in a collision of two Bose-Einstein 
condensates (BECs) of metastable helium~\cite{Perrin:07,Perrin:08,Jaskula:10,Partridge:10}~(see~Fig.~\ref{fig:1}).
The CS inequality which we study is a \textit{multimode} inequality,  
involving integrated atomic densities, and therefore is different from the typical two-mode situation studied in quantum optics. 
Our results demonstrate the potential of atom optics experiments
to extend the fundamental tests of quantum mechanics to ensembles of massive particles.
Indeed, violation of the CS inequality implies the possibility of (but is not equivalent to) formation of 
quantum states that exhibit the Einstein-Podolsky-Rosen (EPR) correlations or violate a 
Bell's inequality \cite{Reid:86}. The EPR and Bell-state correlations are of course of wider significance to foundational principles of 
quantum mechanics than those that violate a CS inequality. Nevertheless, the importance of understanding 
the CS inequality 
in new physical regimes lies in the fact that: (\textit{i}) they are the simplest possible tests of 
stronger-than-classical correlations, and (\textit{ii}) they  can be viewed 
as precursors, or necessary conditions, for the stricter tests of quantum mechanics.

The atom-atom correlations resulting from the collision and violating the CS inequality
are measured after long time-of-flight expansion using time- and position-resolved 
atom detection techniques unique to metastable atoms~\cite{He-Review}. 
The $307$~ms long expansion time combined with a large collision and hence scattering velocity results in a $\sim\!\!6$ cm spatial separation between the scattered, correlated atoms. This separation is quite large compared to what has been achieved in recent related BEC experiments based on double-well or two-component systems
\cite{Esteve:08,Oberthaler-interferometer,Riedel-interferometer}, 
trap modulation techniques \cite{Vienna-twins}, or spin-changing interactions 
\cite{Hannover-twins,Chapman:11}. This makes the BEC collisions  
ideally suited to quantum-nonlocality tests using ultracold atomic gases and 
the intrinsic interatomic interactions.

In a simple two-mode quantum problem,
described by boson creation and annihilation operators $\hat{a}^{\dag}_{i}$ and $\hat{a}_i$ ($i\!=\!1,2$),
the Cauchy-Schwarz inequality of the form of Eq.~(\ref{cs1}) can be formulated in terms of the 
second-order correlation functions, 
{$G^{(2)}_{ij}=\langle:\hat{n}_i \hat{n}_j:\rangle=
\langle \hat{a}^{\dag}_i\hat{a}^{\dag}_j\hat{a}_j\hat{a}_i\rangle$},
and 
reads \cite{Glauber:63,Reid:86,Walls:08}
\begin{equation}
G^{(2)}_{12}\leq [G^{(2)}_{11}G^{(2)}_{22}]^{1/2},
\label{CS-g2-full}
\end{equation}
or simply $G^{(2)}_{12}\!\leq\!G^{(2)}_{11}$ in the symmetric case of $G^{(2)}_{11}=G^{(2)}_{22}$.
Here, $G^{(2)}_{12}=G^{(2)}_{21}$, $\hat{n}_i=\hat{a}^{\dag}_i\hat{a}_i$ is the particle number operator, and the 
double  colons indicate normal ordering of the creation and annihilation operators,
which ensures the correct quantum-mechanical 
interpretation of the process of detection of pairs of particles that contribute to the measurement of the second-order correlation function \cite{Glauber:63}.
Stronger-than-classical correlation violating this inequality would 
require $G^{(2)}_{12}\!>\![G^{(2)}_{11}G^{(2)}_{22}]^{1/2}$, or $G^{(2)}_{12}\!>\!G^{(2)}_{11}$ in the symmetric case.

The situation we analyze here is counterintuitive in that
we observe a peak cross-correlation (for pairs of atoms  scattered in opposite directions) 
that is smaller than the peak auto-correlation 
(for pairs of atoms propagating in the same direction).
In a simple two-mode model such a ratio of the cross-correlation and auto-correlation satisfies
the classical CS inequality. 
However, in order to adequately treat the atom-atom correlations in the BEC collision problem, 
one must generalize the 
CS inequality to a multimode situation, which takes into account the fact that 
the cross- and auto-correlations in matter-wave optics are usually \emph{functions}
(in our case of momentum).
The various correlation functions can have different widths and peak heights, and one must 
define an appropriate integration domain over multiple momentum modes to recover an inequality that
plays the same role as that in the two-mode case and \textit{is} actually violated, as we show below.

The experimental setup was described in Refs.~\cite{Jaskula:10,Partridge:10}.
Briefly, a cigar-shaped BEC of metastable helium, containing approximately $\sim\!\!10^5$ atoms, trapped initially in a harmonic trapping potential with frequencies $(\omega_x, \omega_y, \omega_z)/2\pi=(1500, 1500, 7.5)$ Hz, was split by Bragg diffraction into two parts along the axial ($z$-) direction [see Fig.~\ref{fig:1}(a)],
with velocities differing by twice the single photon recoil velocity $\vrec=9.2$~cm/s.
Atoms interact via binary, momentum conserving $s$-wave collisions and scatter onto a nearly
spherical halo [see Fig.~\ref{fig:1}(b)] whose radius in velocity space is about the recoil velocity \cite{Jaskula:10,Krachmalnicoff:10}. 
The scattered atoms fall onto a detector that records the arrival times 
and positions of individual atoms \cite{He-Review} with a quantum efficiency of $\sim\!10$\%. 
The halo diameter in position space at the detector is $\sim\!6$ cm.
We use the arrival times and positions to reconstruct 3D
velocity vectors $\bf{v}$ for each atom.
The unscattered BECs locally saturate the detector. 
To quantify the strength of correlations corresponding 
only to spontaneously scattered atoms, 
we exclude from the analysis the data points containing 
the BECs and their immediate vicinity ($|v_z|\!<\!0.5\,\vrec$) and further restrict ourselves to a spherical shell of radial thickness $0.9\!<\!v_r/\vrec\!<\!1.1$ (where the signal to noise is large enough), defining the total volume of the analyzed region as $\mathcal{V}_{\mathrm{data}}$.

Using the atom arrival and position data, we can measure the second-order correlation 
functions between the atom number densities $\hat{n}(\mathbf{k})$
at two points in momentum space, ${\cal{G}}^{(2)}(\mathbf{k},\mathbf{k}^{\prime})=\langle:\!\hat{n}(\mathbf{k})\hat{n}(\mathbf{k}^{\prime})\!:\rangle$ (see Supplementary Material \cite{Supplementary}), with $\bf{k}$ denoting 
the wave-vector $\mathbf{k}=m\mathbf{v}/\hbar$ and $\hbar \mathbf{k}$ the momentum.
The correlation measurements are averaged over 
a certain counting zone (integration volume $\mathcal{V}$)
on the scattering sphere in order to get statistically significant results. 
By choosing $\bf{k^{\prime}}$
to be nearly opposite or nearly collinear to $\bf{k}$, 
we can define the averaged back-to-back (BB) or collinear (CL) correlation functions,
\begin{eqnarray}
{\cal{G}}_{\mathrm{BB}}^{(2)}(\Delta \mathbf{k})&=&\int _{\mathcal{V}}d^{3}\mathbf{k}~{\cal{G}}^{(2)}(\mathbf{k},-\mathbf{k}+\Delta \mathbf{k}),\label{g2-BB-av}\\
{\cal{G}}_{\mathrm{CL}}^{(2)}(\Delta \mathbf{k})&=&\int _{\mathcal{V}}d^{3}\mathbf{k}~{\cal{G}}^{(2)}(\mathbf{k},\mathbf{k}+\Delta \mathbf{k}),
\label{g2-CL-av}
\end{eqnarray}
which play a role analogous to the cross- and auto-correlation functions, $G^{(2)}_{12}$
 and $G^{(2)}_{ii}$, in the simple two-mode problem discussed above.
The BB and CL correlations are defined as functions of the relative
displacement $\Delta \mathbf{k}$, while the dependence on $\mathbf{k}$ is lost due to the averaging.

\begin{figure}[tb]
\includegraphics[width=7.5cm]{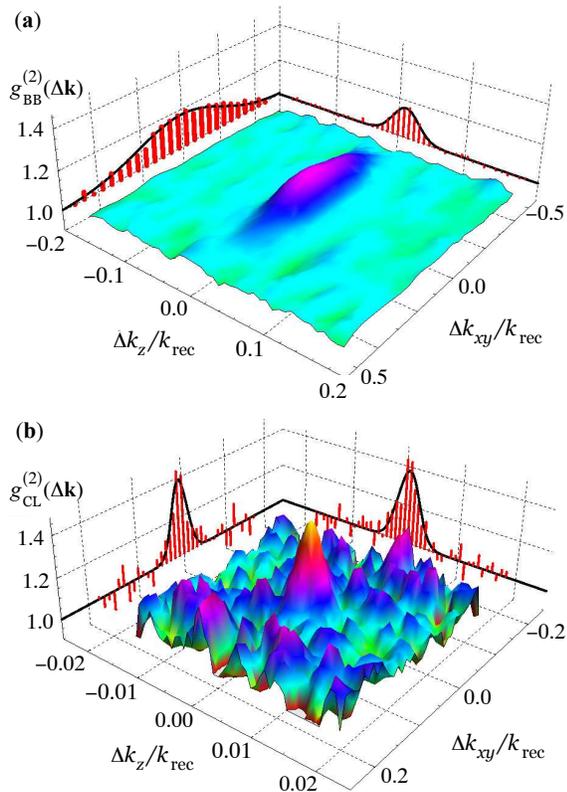}
\caption{(Color online) Normalized back-to-back (a) and collinear (b) correlation functions, $g_{\mathrm{BB}}^{(2)}(\Delta \mathbf{k})$ and $g_{\mathrm{CL}}^{(2)}(\Delta \mathbf{k})$, in momentum space 
integrated over $\mathcal{V}_{\mathrm{data}}$ corresponding to $|k_z|\!<\!0.5\,k_{\mathrm{rec}}$ and $0.9\!<\!k_r/k_{\mathrm{rec}}\!<\!1.1$, where $k_{\mathrm{rec}}\!=\!mv_{\mathrm{rec}}/\hbar$ is the recoil momentum.
The data is averaged over $3600$ experimental runs.
Because of the cylindrical symmetry of the initial condensate and of the overall geometry of the collision, the dependence on the $k_{x}$ and $k_{y}$ components should physically be identical and therefore can be combined (averaged); the correlation functions can then be presented as 2D surface plots on the ($k_z,k_{xy}$) plane. 
The 2D plots were smoothed with a nearest neighbor running average.
The data points along the $k_z$ and $k_{xy}$ projections (corresponding to thin slices centered at $k_{xy}=0$ and $k_z=0$, respectively) are not smoothed.
The solid lines show the Gaussian fits to these projections. 
The peak height of the back-to-back correlation function is $\sim\!1.2$ while that of the collinear correlation function is $\sim\!1.4$, 
apparently confirming the Cauchy-Schwarz inequality. 
The widths of the two distributions are, however, very different ($\sigma_{\mathrm{BB},x}\simeq \sigma_{\mathrm{BB},y}\simeq 0.21k_{\rm{rec}}$, $\sigma_{\mathrm{BB},z}\simeq 0.019 k_{\rm{rec}}$, whereas $\sigma_{\mathrm{CL},x}\simeq \sigma_{\mathrm{CL},y}\simeq 0.036 k_{\rm{rec}}$, $\sigma_{\mathrm{CL},z}\simeq 0.002 k_{\rm{rec}}$)
and a multimode formulation of the Cauchy-Schwarz inequality, 
which relates the relative volumes of the correlation functions,
\emph{is} violated.}
\end{figure}

The normalized BB and CL correlations functions, $g_{\mathrm{BB}}^{(2)}(\Delta \mathbf{k})$ and $g_{\mathrm{CL}}^{(2)}(\Delta \mathbf{k})$, averaged over the unexcised 
part of the scattering sphere $\mathcal{V}_{\mathrm{data}}$ are shown in Fig.~2.
The BB correlation peak results from binary, elastic collisions
between atoms, whereas the CL correlation peak is a variant of the Hanbury
Brown and Twiss effect \cite{Schellekens:05,ANU-He}---a two-particle
interference  involving members of two different atom pairs \cite{Perrin:07,Molmer:08,Perrin:08,Ogren:09}.
Both correlation functions are anisotropic
because of the anisotropy of the initial colliding condensates.

An important difference with the experiment of Ref.~\cite{Perrin:07} is that
the geometry in the present experiment (with vertically elongated condensates) 
is such that the observed widths of the correlation functions are not limited by the detector resolution.
Here we now observe that the BB and CL correlations 
have very different widths, 
with the BB width being significantly larger than the CL width.
This broadening is largely due to the size of the condensate in
the vertical direction ($\sim \!1$~mm).
The elongated nature of the cloud and the estimated temperature of $\sim\!200$ nK also means that the condensates correspond 
in fact to \emph{quasicondensates}~\cite{Petrov:01}
whose phase coherence length is smaller than the size of the atomic cloud.
The broadening of the BB correlation due to the presence of quasicondensates
will be discussed in another paper \cite{InPreparation},
but we emphasize that the CS inequality analyzed here is insensitive to 
the detailed broadening mechanism as it relies on integrals over correlation functions.
This is one of the key points
in considering the multimode CS inequality.

Since the peak of the CL correlation function corresponds to a situation in which the two atoms follow the same path, 
we can associate it with the auto-correlation of the momentum of the particles on the collision sphere. Similarly, the peak of the 
BB correlation function corresponds to two atoms following two distinct paths and therefore can be associated with the 
cross-correlation function between the respective momenta. Hence we realize a situation in which one is tempted to apply the
CS inequality to the peak values of these correlation functions.
As we see from Fig.~2, if one naively uses only the peak heights, 
the CS inequality is \textit{not} violated since $g^{(2)}_{\mathrm{BB}}(0)<g^{(2)}_{\mathrm{CL}}(0)$ and hence ${\cal{G}}_{\mathrm{BB}}^{(2)}(0)<{\cal{G}}_{\mathrm{CL}}^{(2)}(0)$ due to the nearly identical normalization factors \cite{Supplementary}.

We can, however, construct a CS inequality that \emph{is} violated if we use integrated correlation functions, $\overline{\cal{G}}_{ij}^{(2)}$, that correspond 
to atom numbers $\hat{N}_{i}=\int_{\mathcal{V}_{i}}d^{3}\mathbf{k\,}\hat{a}^{\dagger}(\mathbf{k})\mathbf{\,}\hat{a}(\mathbf{k})$ ($i=1,2$) in two distinct zones on the collision halo \cite{Supplementary}: 
\begin{equation}
\overline{\cal{G}}_{ij}^{(2)}=\langle : \hat{N}_i \hat{N}_j : \rangle=\int_{\mathcal{V}_{i}}\!d^{3}\mathbf{k}\int_{\mathcal{V}_{j}}\!d^{3}\mathbf{k}^{\prime}\;{\cal{G}}^{(2)}(\mathbf{k},\mathbf{k}^{\prime}).
\label{eq:final-G}
\end{equation}
The choice of the two integration (zone) volumes $\mathcal{V}_{i}$ and $\mathcal{V}_{j}$ determines whether the $\overline{\cal{G}}_{ij}^{(2)}$-function corresponds to the BB ($i\neq j$) or  CL ($i=j$) correlation functions, Eqs.~(\ref{g2-BB-av}) and (\ref{g2-CL-av}).

The CS inequality that we can now analyze for violation reads $\overline{\cal{G}}^{(2)}_{12}\leq [\overline{\cal{G}}^{(2)}_{11}\overline{\cal{G}}^{(2)}_{22}]^{1/2}$.
To quantify the degree of violation, we introduce a correlation coefficient 
\begin{equation}
C=\overline{\cal{G}}^{(2)}_{12}/[\overline{\cal{G}}^{(2)}_{11}\overline{\cal{G}}^{(2)}_{22}]^{1/2},
\label{eq:C}
\end{equation}
which is smaller than unity classically, but can be larger than unity
for states with stronger-than-classical correlations.

\begin{figure}[tb]
\includegraphics[width=6.0cm]{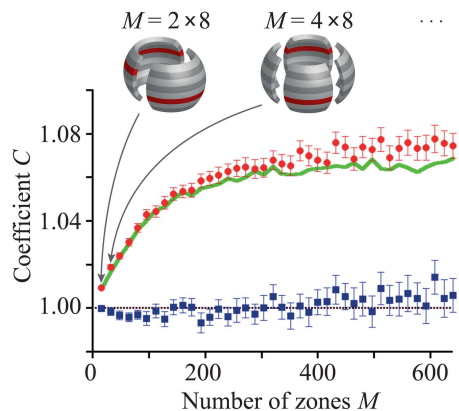}
\caption{(Color online) Correlation coefficient $C$ as a function of the number of zones 
$M=\mathcal{V}_{\mathrm{data}}/\mathcal{V}_{1}$ into which we cut the scattering sphere. 
$C>1$ corresponds to violation of the Cauchy-Schwarz inequality.
The scattering sphere was cut into $8$ polar and from $2$ to $80$ azimuthal zones;
the resulting arrangement of zones for $M\!=\!16$ and $32$ 
is illustrated in the upper panel.
The observed values of $C$ for pairs of correlated diametrically opposite zones (shown in red in the upper panel as an example) were averaged to get one data point for a given $M$; 
the data points for such zones are shown as red circles, for uncorrelated (neighbouring) zones---as blue squares.
The error bars show the standard deviation of the mean over the number of zone pairs. 
The (green) solid curve is the theoretical prediction \cite{Supplementary} calculated using the 
experimental parameters and a stochastic Bogoliubov approach \cite{Krachmalnicoff:10,Deuar:11}.}
\end{figure}

In Fig.~3
we plot the correlation coefficient $C$ determined from the data for different
integration zones $\mathcal{V}_1$ and $\mathcal{V}_2$, but
always keeping the two volumes equal. 
When $\mathcal{V}_1$ and $\mathcal{V}_2$ correspond to diametrically opposed, correlated pairs of zones (red circles), $C$ is greater than 
unity, violating the CS inequality, while for neighboring, uncorrelated pairs (blue squares) the CS inequality is not violated. The figure also shows the results of a quantum-mechanical calculation of $C$ using a 
stochastic Bogoliubov approach (green solid curve) \cite{Krachmalnicoff:10,Deuar:11,Supplementary}.
The calculation is for the initial total number of atoms $N=85\;000$ and is in good agreement with the observations.
The choice of large integration volumes (small number of zones $M$) results in only weak violations, while using
smaller volumes (large $M$) increases the violation. 
This behavior is to be expected \cite{Supplementary} because large integration zones include many, uncorrelated events which dilute the computed correlation.
The saturation of $C$, in the current arrangement of integration zones -- with a fixed number of polar cuts and hence a fixed zone size along $z$ which always remains larger than the longitudinal correlation width -- occurs when the tangential size of the zone begins to approach the
transverse width of the CL correlation function. If the zone sizes were made smaller in all directions, we would recover the situation applicable to the peak values of the correlation functions (and hence no CS violation) as soon as the sizes become smaller than the respective correlations widths (see Eq. (S11) in \cite{Supplementary}).

We have shown the violation of the CS inequality using the experimental data of Ref. \cite{Jaskula:10} in which a sub-Poissonian variance in the atom number difference between opposite zones was observed. Although the two effects are linked mathematically in simple cases, they are not equivalent in general \cite{Marino:08,Supplementary}. 
Because of the multimode nature of the four-wave mixing process, we observe stronger (weaker) suppression of the variance below the shot-noise level for the larger (smaller) zones (see Fig. 3 of \cite{Jaskula:10}), whereas the degree of violation of the CS inequality follows the opposite trend. This difference can be of importance for other experimental tests of stronger-than-classical correlations in inherently multimode situations in matter-wave optics.

The nonclassical character of the observed correlations implies that the scattered atoms cannot be described by classical stochastic random variables \cite{Su-Wodkiewicz:91}. Our experiment is an important step towards the demonstrations of increasingly restrictive types of nonlocal quantum correlations with matter waves, which we hope will one day culminate in the violation of a Bell inequality as well. In this case, the nonclassical character of correlations will also defy a description via a local hidden variable theory \cite{Su-Wodkiewicz:91,Walls:08}. 
Non-optical violations of Bell's inequalities have so far only been demonstrated for \emph{pairs} of massive particles 
(such as two trapped ions \cite{Bell-ions} or proton-proton pairs in the decay of $^2$He \cite{p-p}), but never in the multi-particle regime. 
The BEC collision scheme used here is particularly well-suited  
for demonstrating a 
Bell inequality violation \cite{Lewis-Swan-KK}
using 
an atom optics analog of the Rarity-Tapster setup \cite{Rarity:90}.

We thank P. Zi\'n and T. Wasak for useful discussions. K.V.K. is supported by the ARC FT100100285 grant, P.D. by Polish Government research grants for the years 2010--2013 and the EU contract PERG06-GA-2009-256291, J.R. by the DGA, R.L. by the FCT SFRH/BD/74352/2010, and support for the experimental work comes from the IFRAF program, the Triangle de la Physique, and the ANR grants DESINA and ProQuP.


\begin{thebibliography}{33}%
\makeatletter
\providecommand \@ifxundefined [1]{%
 \@ifx{#1\undefined}
}%
\providecommand \@ifnum [1]{%
 \ifnum #1\expandafter \@firstoftwo
 \else \expandafter \@secondoftwo
 \fi
}%
\providecommand \@ifx [1]{%
 \ifx #1\expandafter \@firstoftwo
 \else \expandafter \@secondoftwo
 \fi
}%
\providecommand \natexlab [1]{#1}%
\providecommand \enquote  [1]{``#1''}%
\providecommand \bibnamefont  [1]{#1}%
\providecommand \bibfnamefont [1]{#1}%
\providecommand \citenamefont [1]{#1}%
\providecommand \href@noop [0]{\@secondoftwo}%
\providecommand \href [0]{\begingroup \@sanitize@url \@href}%
\providecommand \@href[1]{\@@startlink{#1}\@@href}%
\providecommand \@@href[1]{\endgroup#1\@@endlink}%
\providecommand \@sanitize@url [0]{\catcode `\\12\catcode `\$12\catcode
  `\&12\catcode `\#12\catcode `\^12\catcode `\_12\catcode `\%12\relax}%
\providecommand \@@startlink[1]{}%
\providecommand \@@endlink[0]{}%
\providecommand \url  [0]{\begingroup\@sanitize@url \@url }%
\providecommand \@url [1]{\endgroup\@href {#1}{\urlprefix }}%
\providecommand \urlprefix  [0]{URL }%
\providecommand \Eprint [0]{\href }%
\providecommand \doibase [0]{http://dx.doi.org/}%
\providecommand \selectlanguage [0]{\@gobble}%
\providecommand \bibinfo  [0]{\@secondoftwo}%
\providecommand \bibfield  [0]{\@secondoftwo}%
\providecommand \translation [1]{[#1]}%
\providecommand \BibitemOpen [0]{}%
\providecommand \bibitemStop [0]{}%
\providecommand \bibitemNoStop [0]{.\EOS\space}%
\providecommand \EOS [0]{\spacefactor3000\relax}%
\providecommand \BibitemShut  [1]{\csname bibitem#1\endcsname}%
\let\auto@bib@innerbib\@empty
\bibitem [{\citenamefont {Steele}(2004)}]{CS-book}%
  \BibitemOpen
  \bibfield  {author} {\bibinfo {author} {\bibfnamefont {J.~M.}\ \bibnamefont
  {Steele}},\ }\href@noop {} {\emph {\bibinfo {title} {The {C}auchy-{S}chwarz
  Master Class: An Introduction to the Art of Mathematical Inequalities}}}\
  (\bibinfo  {publisher} {Cambridge University Press},\ \bibinfo {address}
  {Cambridge},\ \bibinfo {year} {2004})\BibitemShut {NoStop}%
\bibitem [{\citenamefont {Glauber}(1963)}]{Glauber:63}%
  \BibitemOpen
  \bibfield  {author} {\bibinfo {author} {\bibfnamefont {R.~J.}\ \bibnamefont
  {Glauber}},\ }\href {\doibase 10.1103/PhysRev.130.2529} {\bibfield  {journal}
  {\bibinfo  {journal} {Phys. Rev.}\ }\textbf {\bibinfo {volume} {130}},\
  \bibinfo {pages} {2529} (\bibinfo {year} {1963})}\BibitemShut {NoStop}%
\bibitem [{\citenamefont {Reid}\ and\ \citenamefont {Walls}(1986)}]{Reid:86}%
  \BibitemOpen
  \bibfield  {author} {\bibinfo {author} {\bibfnamefont {M.~D.}\ \bibnamefont
  {Reid}}\ and\ \bibinfo {author} {\bibfnamefont {D.~F.}\ \bibnamefont
  {Walls}},\ }\href {\doibase 10.1103/PhysRevA.34.1260} {\bibfield  {journal}
  {\bibinfo  {journal} {Phys. Rev. A}\ }\textbf {\bibinfo {volume} {34}},\
  \bibinfo {pages} {1260} (\bibinfo {year} {1986})}\BibitemShut {NoStop}%
\bibitem [{\citenamefont {Walls}\ and\ \citenamefont
  {Milburn}(2008)}]{Walls:08}%
  \BibitemOpen
  \bibfield  {author} {\bibinfo {author} {\bibfnamefont {D.~F.}\ \bibnamefont
  {Walls}}\ and\ \bibinfo {author} {\bibfnamefont {G.~J.}\ \bibnamefont
  {Milburn}},\ }\href@noop {} {\emph {\bibinfo {title} {Quantum Optics}}},\
  \bibinfo {edition} {2nd}\ ed.\ (\bibinfo  {publisher} {Springer},\ \bibinfo
  {address} {Berlin},\ \bibinfo {year} {2008})\BibitemShut {NoStop}%
\bibitem [{\citenamefont {Kimble}\ \emph {et~al.}(1977)\citenamefont {Kimble},
  \citenamefont {Dagenais},\ and\ \citenamefont
  {Mandel}}]{Kimble-Dagenais-Mandel-77}%
  \BibitemOpen
  \bibfield  {author} {\bibinfo {author} {\bibfnamefont {H.~J.}\ \bibnamefont
  {Kimble}}, \bibinfo {author} {\bibfnamefont {M.}~\bibnamefont {Dagenais}}, \
  and\ \bibinfo {author} {\bibfnamefont {L.}~\bibnamefont {Mandel}},\ }\href
  {\doibase 10.1103/PhysRevLett.39.691} {\bibfield  {journal} {\bibinfo
  {journal} {Phys. Rev. Lett.}\ }\textbf {\bibinfo {volume} {39}},\ \bibinfo
  {pages} {691} (\bibinfo {year} {1977})}\BibitemShut {NoStop}%
\bibitem [{\citenamefont {Clauser}(1974)}]{Clauser:74}%
  \BibitemOpen
  \bibfield  {author} {\bibinfo {author} {\bibfnamefont {J.~F.}\ \bibnamefont
  {Clauser}},\ }\href@noop {} {\bibfield  {journal} {\bibinfo  {journal} {Phys.
  Rev. D}\ }\textbf {\bibinfo {volume} {9}},\ \bibinfo {pages} {853} (\bibinfo
  {year} {1974})}\BibitemShut {NoStop}%
\bibitem [{\citenamefont {Zou}\ \emph {et~al.}(1991)\citenamefont {Zou},
  \citenamefont {Wang},\ and\ \citenamefont {Mandel}}]{Zou:91}%
  \BibitemOpen
  \bibfield  {author} {\bibinfo {author} {\bibfnamefont {X.}~\bibnamefont
  {Zou}}, \bibinfo {author} {\bibfnamefont {L.~J.}\ \bibnamefont {Wang}}, \
  and\ \bibinfo {author} {\bibfnamefont {L.}~\bibnamefont {Mandel}},\
  }\href@noop {} {\bibfield  {journal} {\bibinfo  {journal} {Opt. Comm.}\
  }\textbf {\bibinfo {volume} {84}},\ \bibinfo {pages} {351} (\bibinfo {year}
  {1991})}\BibitemShut {NoStop}%
\bibitem [{\citenamefont {Marino}\ \emph {et~al.}(2008)\citenamefont {Marino},
  \citenamefont {Boyer},\ and\ \citenamefont {Lett}}]{Marino:08}%
  \BibitemOpen
  \bibfield  {author} {\bibinfo {author} {\bibfnamefont {A.~M.}\ \bibnamefont
  {Marino}}, \bibinfo {author} {\bibfnamefont {V.}~\bibnamefont {Boyer}}, \
  and\ \bibinfo {author} {\bibfnamefont {P.~D.}\ \bibnamefont {Lett}},\ }\href
  {\doibase 10.1103/PhysRevLett.100.233601} {\bibfield  {journal} {\bibinfo
  {journal} {Phys. Rev. Lett.}\ }\textbf {\bibinfo {volume} {100}},\ \bibinfo
  {pages} {233601} (\bibinfo {year} {2008})}\BibitemShut {NoStop}%
\bibitem [{\citenamefont {Perrin}\ \emph {et~al.}(2007)\citenamefont {Perrin},
  \citenamefont {Chang}, \citenamefont {Krachmalnicoff}, \citenamefont
  {Schellekens}, \citenamefont {Boiron}, \citenamefont {Aspect},\ and\
  \citenamefont {Westbrook}}]{Perrin:07}%
  \BibitemOpen
  \bibfield  {author} {\bibinfo {author} {\bibfnamefont {A.}~\bibnamefont
  {Perrin}}, \bibinfo {author} {\bibfnamefont {H.}~\bibnamefont {Chang}},
  \bibinfo {author} {\bibfnamefont {V.}~\bibnamefont {Krachmalnicoff}},
  \bibinfo {author} {\bibfnamefont {M.}~\bibnamefont {Schellekens}}, \bibinfo
  {author} {\bibfnamefont {D.}~\bibnamefont {Boiron}}, \bibinfo {author}
  {\bibfnamefont {A.}~\bibnamefont {Aspect}}, \ and\ \bibinfo {author}
  {\bibfnamefont {C.~I.}\ \bibnamefont {Westbrook}},\ }\href {\doibase
  10.1103/PhysRevLett.99.150405} {\bibfield  {journal} {\bibinfo  {journal}
  {Phys. Rev. Lett.}\ }\textbf {\bibinfo {volume} {99}},\ \bibinfo {pages}
  {150405} (\bibinfo {year} {2007})}\BibitemShut {NoStop}%
\bibitem [{\citenamefont {Perrin}\ \emph {et~al.}(2008)\citenamefont {Perrin},
  \citenamefont {Savage}, \citenamefont {Boiron}, \citenamefont
  {Krachmalnicoff}, \citenamefont {Westbrook},\ and\ \citenamefont
  {Kheruntsyan}}]{Perrin:08}%
  \BibitemOpen
  \bibfield  {author} {\bibinfo {author} {\bibfnamefont {A.}~\bibnamefont
  {Perrin}}, \bibinfo {author} {\bibfnamefont {C.~M.}\ \bibnamefont {Savage}},
  \bibinfo {author} {\bibfnamefont {D.}~\bibnamefont {Boiron}}, \bibinfo
  {author} {\bibfnamefont {V.}~\bibnamefont {Krachmalnicoff}}, \bibinfo
  {author} {\bibfnamefont {C.~I.}\ \bibnamefont {Westbrook}}, \ and\ \bibinfo
  {author} {\bibfnamefont {K.~V.}\ \bibnamefont {Kheruntsyan}},\ }\href@noop {}
  {\bibfield  {journal} {\bibinfo  {journal} {New J. Phys.}\ }\textbf {\bibinfo
  {volume} {10}},\ \bibinfo {pages} {045021} (\bibinfo {year}
  {2008})}\BibitemShut {NoStop}%
\bibitem [{\citenamefont {Jaskula}\ \emph {et~al.}(2010)\citenamefont
  {Jaskula}, \citenamefont {Bonneau}, \citenamefont {Partridge}, \citenamefont
  {Krachmalnicoff}, \citenamefont {Deuar}, \citenamefont {Kheruntsyan},
  \citenamefont {Aspect}, \citenamefont {A.~Boiron},\ and\ \citenamefont
  {Westbrook}}]{Jaskula:10}%
  \BibitemOpen
  \bibfield  {author} {\bibinfo {author} {\bibfnamefont {J.-C.}\ \bibnamefont
  {Jaskula}}, \bibinfo {author} {\bibfnamefont {M.}~\bibnamefont {Bonneau}},
  \bibinfo {author} {\bibfnamefont {G.~B.}\ \bibnamefont {Partridge}}, \bibinfo
  {author} {\bibfnamefont {V.}~\bibnamefont {Krachmalnicoff}}, \bibinfo
  {author} {\bibfnamefont {P.}~\bibnamefont {Deuar}}, \bibinfo {author}
  {\bibfnamefont {K.~V.}\ \bibnamefont {Kheruntsyan}}, \bibinfo {author}
  {\bibnamefont {Aspect}}, \bibinfo {author} {\bibfnamefont {D.}~\bibnamefont
  {A.~Boiron}}, \ and\ \bibinfo {author} {\bibfnamefont {C.~I.}\ \bibnamefont
  {Westbrook}},\ }\href {\doibase 10.1103/PhysRevLett.104.190402} {\bibfield
  {journal} {\bibinfo  {journal} {Phys. Rev. Lett.}\ }\textbf {\bibinfo
  {volume} {105}},\ \bibinfo {pages} {190402} (\bibinfo {year}
  {2010})}\BibitemShut {NoStop}%
\bibitem [{\citenamefont {Partridge}\ \emph {et~al.}(2010)\citenamefont
  {Partridge}, \citenamefont {Jaskula}, \citenamefont {Bonneau}, \citenamefont
  {Boiron},\ and\ \citenamefont {Westbrook}}]{Partridge:10}%
  \BibitemOpen
  \bibfield  {author} {\bibinfo {author} {\bibfnamefont {G.~B.}\ \bibnamefont
  {Partridge}}, \bibinfo {author} {\bibfnamefont {J.-C.}\ \bibnamefont
  {Jaskula}}, \bibinfo {author} {\bibfnamefont {M.}~\bibnamefont {Bonneau}},
  \bibinfo {author} {\bibfnamefont {D.}~\bibnamefont {Boiron}}, \ and\ \bibinfo
  {author} {\bibfnamefont {C.~I.}\ \bibnamefont {Westbrook}},\ }\href {\doibase
  http://link.aps.org/doi/10.1103/PhysRevA.81.053631} {\bibfield  {journal}
  {\bibinfo  {journal} {Phys. Rev. A}\ }\textbf {\bibinfo {volume} {81}},\
  \bibinfo {pages} {053631} (\bibinfo {year} {2010})}\BibitemShut {NoStop}%
\bibitem [{\citenamefont {Vassen}\ \emph {et~al.}(2012)\citenamefont {Vassen},
  \citenamefont {Cohen-Tannoudji}, \citenamefont {Leduc}, \citenamefont
  {Boiron}, \citenamefont {Westbrook}, \citenamefont {Truscott}, \citenamefont
  {Baldwin}, \citenamefont {Birkl}, \citenamefont {Cancio},\ and\ \citenamefont
  {Trippenbach}}]{He-Review}%
  \BibitemOpen
  \bibfield  {author} {\bibinfo {author} {\bibfnamefont {W.}~\bibnamefont
  {Vassen}}, \bibinfo {author} {\bibfnamefont {C.}~\bibnamefont
  {Cohen-Tannoudji}}, \bibinfo {author} {\bibfnamefont {M.}~\bibnamefont
  {Leduc}}, \bibinfo {author} {\bibfnamefont {D.}~\bibnamefont {Boiron}},
  \bibinfo {author} {\bibfnamefont {C.~I.}\ \bibnamefont {Westbrook}}, \bibinfo
  {author} {\bibfnamefont {A.}~\bibnamefont {Truscott}}, \bibinfo {author}
  {\bibfnamefont {K.}~\bibnamefont {Baldwin}}, \bibinfo {author} {\bibfnamefont
  {G.}~\bibnamefont {Birkl}}, \bibinfo {author} {\bibfnamefont
  {P.}~\bibnamefont {Cancio}}, \ and\ \bibinfo {author} {\bibfnamefont
  {M.}~\bibnamefont {Trippenbach}},\ }\href {\doibase
  10.1103/RevModPhys.84.175} {\bibfield  {journal} {\bibinfo  {journal} {Rev.
  Mod. Phys.}\ }\textbf {\bibinfo {volume} {84}},\ \bibinfo {pages} {175}
  (\bibinfo {year} {2012})}\BibitemShut {NoStop}%
\bibitem [{\citenamefont {Est\`{e}ve}\ \emph {et~al.}(2008)\citenamefont
  {Est\`{e}ve}, \citenamefont {Gross}, \citenamefont {Weller}, \citenamefont
  {Giovanazzi},\ and\ \citenamefont {Oberthaler}}]{Esteve:08}%
  \BibitemOpen
  \bibfield  {author} {\bibinfo {author} {\bibfnamefont {J.}~\bibnamefont
  {Est\`{e}ve}}, \bibinfo {author} {\bibfnamefont {C.}~\bibnamefont {Gross}},
  \bibinfo {author} {\bibfnamefont {A.}~\bibnamefont {Weller}}, \bibinfo
  {author} {\bibfnamefont {S.}~\bibnamefont {Giovanazzi}}, \ and\ \bibinfo
  {author} {\bibfnamefont {M.~K.}\ \bibnamefont {Oberthaler}},\ }\href@noop {}
  {\bibfield  {journal} {\bibinfo  {journal} {Nature}\ }\textbf {\bibinfo
  {volume} {455}},\ \bibinfo {pages} {1216} (\bibinfo {year}
  {2008})}\BibitemShut {NoStop}%
\bibitem [{\citenamefont {Gross}\ \emph {et~al.}(2010)\citenamefont {Gross},
  \citenamefont {Zibold}, \citenamefont {Nicklas}, \citenamefont {Est\`eve},\
  and\ \citenamefont {Oberthaler}}]{Oberthaler-interferometer}%
  \BibitemOpen
  \bibfield  {author} {\bibinfo {author} {\bibfnamefont {C.}~\bibnamefont
  {Gross}}, \bibinfo {author} {\bibfnamefont {T.}~\bibnamefont {Zibold}},
  \bibinfo {author} {\bibfnamefont {E.}~\bibnamefont {Nicklas}}, \bibinfo
  {author} {\bibfnamefont {J.}~\bibnamefont {Est\`eve}}, \ and\ \bibinfo
  {author} {\bibfnamefont {M.~K.}\ \bibnamefont {Oberthaler}},\ }\href@noop {}
  {\bibfield  {journal} {\bibinfo  {journal} {{Nature}}\ }\textbf {\bibinfo
  {volume} {464}},\ \bibinfo {pages} {1165} (\bibinfo {year}
  {2010})}\BibitemShut {NoStop}%
\bibitem [{\citenamefont {Riedel}\ \emph {et~al.}(2010)\citenamefont {Riedel},
  \citenamefont {B\"ohi}, \citenamefont {Li}, \citenamefont {H\"ansch},
  \citenamefont {Sinatra},\ and\ \citenamefont
  {Treutlein}}]{Riedel-interferometer}%
  \BibitemOpen
  \bibfield  {author} {\bibinfo {author} {\bibfnamefont {M.~F.}\ \bibnamefont
  {Riedel}}, \bibinfo {author} {\bibfnamefont {P.}~\bibnamefont {B\"ohi}},
  \bibinfo {author} {\bibfnamefont {Y.}~\bibnamefont {Li}}, \bibinfo {author}
  {\bibfnamefont {T.~W.}\ \bibnamefont {H\"ansch}}, \bibinfo {author}
  {\bibfnamefont {A.}~\bibnamefont {Sinatra}}, \ and\ \bibinfo {author}
  {\bibfnamefont {P.}~\bibnamefont {Treutlein}},\ }\href@noop {} {\bibfield
  {journal} {\bibinfo  {journal} {{Nature}}\ }\textbf {\bibinfo {volume}
  {464}},\ \bibinfo {pages} {1170} (\bibinfo {year} {2010})}\BibitemShut
  {NoStop}%
\bibitem [{\citenamefont {B\"ucker}\ \emph {et~al.}(2011)\citenamefont
  {B\"ucker}, \citenamefont {Grond}, \citenamefont {Manz}, \citenamefont
  {Berrada}, \citenamefont {Betz}, \citenamefont {Koller}, \citenamefont
  {Hohenester}, \citenamefont {Schumm}, \citenamefont {Perrin},\ and\
  \citenamefont {Schmiedmayer}}]{Vienna-twins}%
  \BibitemOpen
  \bibfield  {author} {\bibinfo {author} {\bibfnamefont {R.}~\bibnamefont
  {B\"ucker}}, \bibinfo {author} {\bibfnamefont {J.}~\bibnamefont {Grond}},
  \bibinfo {author} {\bibfnamefont {S.}~\bibnamefont {Manz}}, \bibinfo {author}
  {\bibfnamefont {T.}~\bibnamefont {Berrada}}, \bibinfo {author} {\bibfnamefont
  {T.}~\bibnamefont {Betz}}, \bibinfo {author} {\bibfnamefont {C.}~\bibnamefont
  {Koller}}, \bibinfo {author} {\bibfnamefont {U.}~\bibnamefont {Hohenester}},
  \bibinfo {author} {\bibfnamefont {T.}~\bibnamefont {Schumm}}, \bibinfo
  {author} {\bibfnamefont {A.}~\bibnamefont {Perrin}}, \ and\ \bibinfo {author}
  {\bibfnamefont {J.}~\bibnamefont {Schmiedmayer}},\ }\href@noop {} {\bibfield
  {journal} {\bibinfo  {journal} {Nature Physics}\ }\textbf {\bibinfo {volume}
  {7}},\ \bibinfo {pages} {608} (\bibinfo {year} {2011})}\BibitemShut {NoStop}%
\bibitem [{\citenamefont {L\"ucke}\ \emph {et~al.}(2011)\citenamefont
  {L\"ucke}, \citenamefont {Scherer}, \citenamefont {Kruse}, \citenamefont
  {Pezz\'e}, \citenamefont {Deuretzbacher}, \citenamefont {Hyllus},
  \citenamefont {Topic}, \citenamefont {Peise}, \citenamefont {Ertmer},
  \citenamefont {Arlt}, \citenamefont {Santos}, \citenamefont {Smerzi},\ and\
  \citenamefont {Klempt}}]{Hannover-twins}%
  \BibitemOpen
  \bibfield  {author} {\bibinfo {author} {\bibfnamefont {B.}~\bibnamefont
  {L\"ucke}}, \bibinfo {author} {\bibfnamefont {M.}~\bibnamefont {Scherer}},
  \bibinfo {author} {\bibfnamefont {J.}~\bibnamefont {Kruse}}, \bibinfo
  {author} {\bibfnamefont {L.}~\bibnamefont {Pezz\'e}}, \bibinfo {author}
  {\bibfnamefont {F.}~\bibnamefont {Deuretzbacher}}, \bibinfo {author}
  {\bibfnamefont {P.}~\bibnamefont {Hyllus}}, \bibinfo {author} {\bibfnamefont
  {O.}~\bibnamefont {Topic}}, \bibinfo {author} {\bibfnamefont
  {J.}~\bibnamefont {Peise}}, \bibinfo {author} {\bibfnamefont
  {W.}~\bibnamefont {Ertmer}}, \bibinfo {author} {\bibfnamefont
  {J.}~\bibnamefont {Arlt}}, \bibinfo {author} {\bibfnamefont {L.}~\bibnamefont
  {Santos}}, \bibinfo {author} {\bibfnamefont {A.}~\bibnamefont {Smerzi}}, \
  and\ \bibinfo {author} {\bibfnamefont {C.}~\bibnamefont {Klempt}},\
  }\href@noop {} {\bibfield  {journal} {\bibinfo  {journal} {Science}\ }\textbf
  {\bibinfo {volume} {334}},\ \bibinfo {pages} {773} (\bibinfo {year}
  {2011})}\BibitemShut {NoStop}%
\bibitem [{\citenamefont {Bookjans}\ \emph {et~al.}(2011)\citenamefont
  {Bookjans}, \citenamefont {Hamley},\ and\ \citenamefont
  {Chapman}}]{Chapman:11}%
  \BibitemOpen
  \bibfield  {author} {\bibinfo {author} {\bibfnamefont {E.~M.}\ \bibnamefont
  {Bookjans}}, \bibinfo {author} {\bibfnamefont {C.~D.}\ \bibnamefont
  {Hamley}}, \ and\ \bibinfo {author} {\bibfnamefont {M.~S.}\ \bibnamefont
  {Chapman}},\ }\href {\doibase 10.1103/PhysRevLett.107.210406} {\bibfield
  {journal} {\bibinfo  {journal} {Phys. Rev. Lett.}\ }\textbf {\bibinfo
  {volume} {107}},\ \bibinfo {pages} {210406} (\bibinfo {year}
  {2011})}\BibitemShut {NoStop}%
\bibitem [{\citenamefont {Krachmalnicoff}\ \emph {et~al.}(2010)\citenamefont
  {Krachmalnicoff}, \citenamefont {Jaskula}, \citenamefont {Bonneau},
  \citenamefont {Leung}, \citenamefont {Partridge}, \citenamefont {Boiron},
  \citenamefont {Westbrook}, \citenamefont {Deuar}, \citenamefont
  {Zi\ifmmode~\acute{n}\else \'{n}\fi{}}, \citenamefont {Trippenbach},\ and\
  \citenamefont {Kheruntsyan}}]{Krachmalnicoff:10}%
  \BibitemOpen
  \bibfield  {author} {\bibinfo {author} {\bibfnamefont {V.}~\bibnamefont
  {Krachmalnicoff}}, \bibinfo {author} {\bibfnamefont {J.-C.}\ \bibnamefont
  {Jaskula}}, \bibinfo {author} {\bibfnamefont {M.}~\bibnamefont {Bonneau}},
  \bibinfo {author} {\bibfnamefont {V.}~\bibnamefont {Leung}}, \bibinfo
  {author} {\bibfnamefont {G.~B.}\ \bibnamefont {Partridge}}, \bibinfo {author}
  {\bibfnamefont {D.}~\bibnamefont {Boiron}}, \bibinfo {author} {\bibfnamefont
  {C.~I.}\ \bibnamefont {Westbrook}}, \bibinfo {author} {\bibfnamefont
  {P.}~\bibnamefont {Deuar}}, \bibinfo {author} {\bibfnamefont
  {P.}~\bibnamefont {Zi\ifmmode~\acute{n}\else \'{n}\fi{}}}, \bibinfo {author}
  {\bibfnamefont {M.}~\bibnamefont {Trippenbach}}, \ and\ \bibinfo {author}
  {\bibfnamefont {K.~V.}\ \bibnamefont {Kheruntsyan}},\ }\href {\doibase
  10.1103/PhysRevLett.104.150402} {\bibfield  {journal} {\bibinfo  {journal}
  {Phys. Rev. Lett.}\ }\textbf {\bibinfo {volume} {104}},\ \bibinfo {pages}
  {150402} (\bibinfo {year} {2010})}\BibitemShut {NoStop}%
\bibitem [{Sup()}]{Supplementary}%
  \BibitemOpen
  \href@noop {} {}\bibinfo {note} {See Supplementary Material at
  http://link.aps.org/ supplemental/10.1103/PhysRevLett.108.260401 for a more
  detailed discussion of correlation functions, the correlation coefficient C,
  and simulation methods.}\BibitemShut {Stop}%
\bibitem [{\citenamefont {Schellekens}\ \emph {et~al.}(2005)\citenamefont
  {Schellekens}, \citenamefont {Hoppeler}, \citenamefont {Perrin},
  \citenamefont {Viana~Gomes}, \citenamefont {Boiron}, \citenamefont {Aspect},\
  and\ \citenamefont {Westbrook}}]{Schellekens:05}%
  \BibitemOpen
  \bibfield  {author} {\bibinfo {author} {\bibfnamefont {M.}~\bibnamefont
  {Schellekens}}, \bibinfo {author} {\bibfnamefont {R.}~\bibnamefont
  {Hoppeler}}, \bibinfo {author} {\bibfnamefont {A.}~\bibnamefont {Perrin}},
  \bibinfo {author} {\bibfnamefont {J.}~\bibnamefont {Viana~Gomes}}, \bibinfo
  {author} {\bibfnamefont {D.}~\bibnamefont {Boiron}}, \bibinfo {author}
  {\bibfnamefont {A.}~\bibnamefont {Aspect}}, \ and\ \bibinfo {author}
  {\bibfnamefont {C.~I.}\ \bibnamefont {Westbrook}},\ }\href@noop {} {\bibfield
   {journal} {\bibinfo  {journal} {Science}\ }\textbf {\bibinfo {volume}
  {310}},\ \bibinfo {pages} {648} (\bibinfo {year} {2005})}\BibitemShut
  {NoStop}%
\bibitem [{\citenamefont {Hodgman}\ \emph {et~al.}(2011)\citenamefont
  {Hodgman}, \citenamefont {Dall}, \citenamefont {Manning}, \citenamefont
  {Baldwin},\ and\ \citenamefont {Truscott}}]{ANU-He}%
  \BibitemOpen
  \bibfield  {author} {\bibinfo {author} {\bibfnamefont {S.~S.}\ \bibnamefont
  {Hodgman}}, \bibinfo {author} {\bibfnamefont {R.~G.}\ \bibnamefont {Dall}},
  \bibinfo {author} {\bibfnamefont {A.~G.}\ \bibnamefont {Manning}}, \bibinfo
  {author} {\bibfnamefont {K.~G.~H.}\ \bibnamefont {Baldwin}}, \ and\ \bibinfo
  {author} {\bibfnamefont {A.~G.}\ \bibnamefont {Truscott}},\ }\href@noop {}
  {\bibfield  {journal} {\bibinfo  {journal} {Science}\ }\textbf {\bibinfo
  {volume} {331}},\ \bibinfo {pages} {1046} (\bibinfo {year}
  {2011})}\BibitemShut {NoStop}%
\bibitem [{\citenamefont {M{\o}lmer}\ \emph {et~al.}(2008)\citenamefont
  {M{\o}lmer}, \citenamefont {Perrin}, \citenamefont {Krachmalnicoff},
  \citenamefont {Leung}, \citenamefont {Boiron}, \citenamefont {Aspect},\ and\
  \citenamefont {Westbrook}}]{Molmer:08}%
  \BibitemOpen
  \bibfield  {author} {\bibinfo {author} {\bibfnamefont {K.}~\bibnamefont
  {M{\o}lmer}}, \bibinfo {author} {\bibfnamefont {A.}~\bibnamefont {Perrin}},
  \bibinfo {author} {\bibfnamefont {V.}~\bibnamefont {Krachmalnicoff}},
  \bibinfo {author} {\bibfnamefont {V.}~\bibnamefont {Leung}}, \bibinfo
  {author} {\bibfnamefont {D.}~\bibnamefont {Boiron}}, \bibinfo {author}
  {\bibfnamefont {A.}~\bibnamefont {Aspect}}, \ and\ \bibinfo {author}
  {\bibfnamefont {C.~I.}\ \bibnamefont {Westbrook}},\ }\href@noop {} {\bibfield
   {journal} {\bibinfo  {journal} {Phys. Rev. A}\ }\textbf {\bibinfo {volume}
  {77}},\ \bibinfo {pages} {033601} (\bibinfo {year} {2008})}\BibitemShut
  {NoStop}%
\bibitem [{\citenamefont {\"Ogren}\ and\ \citenamefont
  {Kheruntsyan}(2009)}]{Ogren:09}%
  \BibitemOpen
  \bibfield  {author} {\bibinfo {author} {\bibfnamefont {M.}~\bibnamefont
  {\"Ogren}}\ and\ \bibinfo {author} {\bibfnamefont {K.~V.}\ \bibnamefont
  {Kheruntsyan}},\ }\href {\doibase 10.1103/PhysRevA.79.021606} {\bibfield
  {journal} {\bibinfo  {journal} {{Phys. Rev. A}}\ }\textbf {\bibinfo {volume}
  {79}},\ \bibinfo {pages} {021606} (\bibinfo {year} {2009})}\BibitemShut
  {NoStop}%
\bibitem [{\citenamefont {Petrov}\ \emph {et~al.}(2001)\citenamefont {Petrov},
  \citenamefont {Shlyapnikov},\ and\ \citenamefont {Walraven}}]{Petrov:01}%
  \BibitemOpen
  \bibfield  {author} {\bibinfo {author} {\bibfnamefont {D.~S.}\ \bibnamefont
  {Petrov}}, \bibinfo {author} {\bibfnamefont {G.~V.}\ \bibnamefont
  {Shlyapnikov}}, \ and\ \bibinfo {author} {\bibfnamefont {J.~T.~M.}\
  \bibnamefont {Walraven}},\ }\href {\doibase 10.1103/PhysRevLett.87.050404}
  {\bibfield  {journal} {\bibinfo  {journal} {Phys. Rev. Lett.}\ }\textbf
  {\bibinfo {volume} {87}},\ \bibinfo {pages} {050404} (\bibinfo {year}
  {2001})}\BibitemShut {NoStop}%
\bibitem [{InP()}]{InPreparation}%
  \BibitemOpen
  \href@noop {} {}\bibinfo {note} {P. Deuar \textit{et al.}, in
  preparation.}\BibitemShut {Stop}%
\bibitem [{\citenamefont {Deuar}\ \emph {et~al.}(2011)\citenamefont {Deuar},
  \citenamefont {Chwede\'nczuk}, \citenamefont {Trippenbach},\ and\
  \citenamefont {Zi\'n}}]{Deuar:11}%
  \BibitemOpen
  \bibfield  {author} {\bibinfo {author} {\bibfnamefont {P.}~\bibnamefont
  {Deuar}}, \bibinfo {author} {\bibfnamefont {J.}~\bibnamefont
  {Chwede\'nczuk}}, \bibinfo {author} {\bibfnamefont {M.}~\bibnamefont
  {Trippenbach}}, \ and\ \bibinfo {author} {\bibfnamefont {P.}~\bibnamefont
  {Zi\'n}},\ }\href {\doibase 10.1103/PhysRevA.83.063625} {\bibfield  {journal}
  {\bibinfo  {journal} {Phys. Rev. A}\ }\textbf {\bibinfo {volume} {83}},\
  \bibinfo {pages} {063625} (\bibinfo {year} {2011})}\BibitemShut {NoStop}%
\bibitem [{\citenamefont {Su}\ and\ \citenamefont
  {W\'odkiewicz}(1991)}]{Su-Wodkiewicz:91}%
  \BibitemOpen
  \bibfield  {author} {\bibinfo {author} {\bibfnamefont {C.}~\bibnamefont
  {Su}}\ and\ \bibinfo {author} {\bibfnamefont {K.}~\bibnamefont
  {W\'odkiewicz}},\ }\href {\doibase
  http://link.aps.org/doi/10.1103/PhysRevA.44.6097} {\bibfield  {journal}
  {\bibinfo  {journal} {Phys. Rev. A}\ }\textbf {\bibinfo {volume} {44}},\
  \bibinfo {pages} {6097} (\bibinfo {year} {1991})}\BibitemShut {NoStop}%
\bibitem [{\citenamefont {Rowe}\ \emph {et~al.}(2001)\citenamefont {Rowe},
  \citenamefont {Kielpinski}, \citenamefont {Meyer}, \citenamefont {Sackett},
  \citenamefont {Itano}, \citenamefont {Monroe},\ and\ \citenamefont
  {Wineland}}]{Bell-ions}%
  \BibitemOpen
  \bibfield  {author} {\bibinfo {author} {\bibfnamefont {M.~A.}\ \bibnamefont
  {Rowe}}, \bibinfo {author} {\bibfnamefont {D.}~\bibnamefont {Kielpinski}},
  \bibinfo {author} {\bibfnamefont {V.}~\bibnamefont {Meyer}}, \bibinfo
  {author} {\bibfnamefont {C.~A.}\ \bibnamefont {Sackett}}, \bibinfo {author}
  {\bibfnamefont {W.~M.}\ \bibnamefont {Itano}}, \bibinfo {author}
  {\bibfnamefont {C.}~\bibnamefont {Monroe}}, \ and\ \bibinfo {author}
  {\bibfnamefont {D.~J.}\ \bibnamefont {Wineland}},\ }\href@noop {} {\bibfield
  {journal} {\bibinfo  {journal} {{Nature}}\ }\textbf {\bibinfo {volume}
  {409}},\ \bibinfo {pages} {791} (\bibinfo {year} {2001})}\BibitemShut
  {NoStop}%
\bibitem [{\citenamefont {Sakai}\ \emph {et~al.}(2006)\citenamefont {Sakai}
  \emph {et~al.}}]{p-p}%
  \BibitemOpen
  \bibfield  {author} {\bibinfo {author} {\bibfnamefont {H.}~\bibnamefont
  {Sakai}} \emph {et~al.},\ }\href {\doibase 10.1103/PhysRevLett.97.150405}
  {\bibfield  {journal} {\bibinfo  {journal} {Phys. Rev. Lett.}\ }\textbf
  {\bibinfo {volume} {97}},\ \bibinfo {pages} {150405} (\bibinfo {year}
  {2006})}\BibitemShut {NoStop}%
\bibitem [{Lew()}]{Lewis-Swan-KK}%
  \BibitemOpen
  \href@noop {} {}\bibinfo {note} {R. Lewis-Swan and K. V. Kheruntsyan (to be
  published).}\BibitemShut {Stop}%
\bibitem [{\citenamefont {Rarity}\ and\ \citenamefont
  {Tapster}(1990)}]{Rarity:90}%
  \BibitemOpen
  \bibfield  {author} {\bibinfo {author} {\bibfnamefont {J.~G.}\ \bibnamefont
  {Rarity}}\ and\ \bibinfo {author} {\bibfnamefont {P.~R.}\ \bibnamefont
  {Tapster}},\ }\href {\doibase 10.1103/PhysRevLett.64.2495} {\bibfield
  {journal} {\bibinfo  {journal} {Phys. Rev. Lett.}\ }\textbf {\bibinfo
  {volume} {64}},\ \bibinfo {pages} {2495} (\bibinfo {year}
  {1990})}\BibitemShut {NoStop}%
\end{thebibliography}
\end{document}